\documentclass{jfm}

\usepackage{graphicx}
\usepackage[round]{natbib}
\usepackage{color}

\ifCUPmtlplainloaded \else
  \checkfont{eurm10}
  \iffontfound
    \IfFileExists{upmath.sty}
      {\typeout{^^JFound AMS Euler Roman fonts on the system,
                   using the 'upmath' package.^^J}%
       \usepackage{upmath}}
      {\typeout{^^JFound AMS Euler Roman fonts on the system, but you
                   dont seem to have the}%
       \typeout{'upmath' package installed. JFM.cls can take advantage
                 of these fonts,^^Jif you use 'upmath' package.^^J}%
      }
  \else
  \fi
\fi

\ifCUPmtlplainloaded \else
  \checkfont{msam10}
  \iffontfound
    \IfFileExists{amssymb.sty}
      {\typeout{^^JFound AMS Symbol fonts on the system, using the
                'amssymb' package.^^J}%
       \usepackage{amssymb}%

      }{}
  \fi
\fi

\ifCUPmtlplainloaded \else
  \IfFileExists{amsbsy.sty}
    {\typeout{^^JFound the 'amsbsy' package on the system, using it.^^J}%
     \usepackage{amsbsy}}
    {}
\fi

\newsavebox{\astrutbox}
\sbox{\astrutbox}{\rule[-5pt]{0pt}{20pt}}

\newcommand{\tabincell}[2]{\begin{tabular}{@{}#1@{}}#2\end{tabular}}

\title[Dynamics and flow-coupling in two-layer turbulent thermal convection]{Dynamics and flow-coupling in two-layer turbulent thermal convection}

\author[Y.-C. Xie and K.-Q. Xia]%
{Y\ls i\ls-C\ls h\ls a\ls o\ns X\ls i\ls e \and \ls K\ls e\ls -\ls Q\ls i\ls n\ls g\ns X\ls i\ls a\ls
\thanks{Email address for correspondence: kxia@phy.cuhk.edu.hk}\ns}
\affiliation{Department of Physics, The Chinese University of Hong Kong, Shatin, Hong Kong, China}

\pubyear{2010}
\volume{650}
\pagerange{119--126}
\date{?; revised ?; accepted ?. - To be entered by editorial office}
\begin{document}

\maketitle

\begin{abstract}
We present an experimental investigation of the dynamics and flow-coupling of convective turbulent flows in a cylindrical Rayleigh-B\'enard convection (RBC) cell with two immiscible fluids, water and fluorinert FC-77 electronic liquid (FC77). With the lighter water above FC77, the latter is under the condition of constant heat flux at its top and bottom boundaries. It is found that one large-scale circulation (LSC) roll exists in each of the fluid layers, and that their circulation planes have two preferred azimuthal orientations separated by $\sim\pi$. A surprising finding of the study is that cessations/reversals of the LSC in FC77 of the two-layer system occur much more frequently than they do in single-layer turbulent RBC, and that a cessation is most likely to result in a flow reversal of the LSC, which is in sharp contrast with the uniform distribution of the orientational angular change of the LSC before and after cessations in single-layer turbulent RBC. This implies that the dynamics governing cessations and reversals in the two systems are very different. Two coupling modes, thermal coupling (flow directions of the two LSCs are opposite to each other at the fluid-fluid interface) and viscous coupling (flow directions of the two LSCs are the same at the fluid-fluid interface), are identified with the former one as the predominant mode. That most cessations (in the FC77 layer) end up as reversals can be understood as a symmetry breaking imposed by the orientation of the LSC in the water layer, which remained unchanged most of the time. Furthermore, the frequently occurring cessations and reversals are caused by the system switching between its two metastable states, i.e. thermal and viscous coupling modes. It is also observed that the strength of the LSC in water becomes \textit{weaker} when the LSC in FC77 rotates faster azimuthally and that the flow strength in FC77 becomes \textit{stronger}
when the LSC in water rotates faster azimuthally, i.e. the influence of the LSC in one fluid layer on the other is not symmetric.

\end{abstract}


\section{Introduction}\label{sec:introduction}
Multilayer turbulent flow, in particular, multilayer turbulent convection is a phenomenon occurring widely in nature. An example is the coupled system of atmospheric and oceanic convections, in which convection is either thermally driven or density driven. Other examples of convection involving coupled fluid layers are the earth's upper and lower mantle \citep{Richter1974}, and the lower mantle and the outer core. For industrial applications, multilayer thermal convection is associated with the liquid encapsulated crystal growth technique \citep{Prakash1994383} and understanding this particular phenomenon will help to improve crystal growth. 
    
Rayleigh-B\'{e}nard convection (RBC), a fluid layer confined between two horizontally parallel plates cooled from above and heated from below, has been studied extensively as an idealised model for turbulent thermal convection \citep*[for reviews, see][] {ahlers2009RMP,Xia2010ARFM,SchumacherChilla2012}. The system is controlled by three parameters, namely the Rayleigh number $Ra=\alpha g \Delta T H^{3}/ (\nu\kappa)$, the Prandtl number $Pr=\nu / \kappa$ and the aspect ratio $\Gamma=D/H$, where $g$ is the gravitational acceleration, $\Delta T$ the temperature difference across a fluid layer of height $H$, $D$ the diameter of the convection cell, $\alpha$, $\nu$ and $\kappa$ are  respectively the thermal expansion coefficient, kinematic viscosity and the thermal diffusivity of the convecting fluid. One of the fascinating features in turbulent RBC is the emergence of a coherent large-scale circulation (LSC), which results from the clustering and self-organization of thermal plumes erupted from the top and bottom thermal boundary layers \citep*{Xi2004JFM}. Ever since the discovery of the LSC by \citet*{Howard1981}, extensive experimental studies of it's structure and dynamics have been made in single-layer turbulent thermal convection \citep[see, e.g. ][]{Cioni1997JFM,Niemela2001JFM,Qiu2001PRE,Sreeni2002PRE,Roche2002EPL,Xia2003PRE,Ahlers2004PRL,Chilla2004EPJB,Thess2004PRE,Eric2005PRL,CSun2005PRL,S2005PRE,hdxi2006PRE,Resagk2006PoF,BA06JFM,hdxi2007PRE,hdxi2008PRE,hdxi2008POF,hdxi2009PRL,qzhou2009JFM,AhlersJFM2009,SA2011JFM,Xie2013JFM}.	

Multilayer RBC system is similar to  traditional single-layer RBC except that multilayer immiscible fluids are used as the working fluids \citep[see, e.g. ][]{Nataf1988FPF,Prakash1994383,Buss2009PRE}. So far, most of multilayer RBC studies are conducted in the non-turbulent regime. Like in the single-layer case, $Ra,~Pr$ and $\Gamma$ remain as the control parameters for convection in each of the fluid layers. As most of convections involving multilayer fluids in nature are in  turbulent regime, it is highly desirable that such phenomenon be investigated in a laboratory setting, where experimental parameters can be precisely controlled. Owing to the presence of a fluid-fluid interface, the boundary conditions of the fluid layers confined between the top and bottom plates will be different from the traditional RBC. In two-layer turbulent RBC, the temperature boundary conditions of the top layer are constant temperature at its top and constant heat flux at its bottom, whereas those of the bottom layer are constant heat flux at both its top and bottom. Besides boundary conditions, the flow in one layer will couple with that in the other and results in changes of the global responses of the system, i.e. heat transport efficiency and dynamics of flow in each of the fluid layer. How do the large-scale flows couple and interact with each other and result in different dynamics of the system are the questions we would like to address in the present study.

    The organization of the paper is as follows. The experimental setup and data analysis methods are introduced in \S~\ref{sec:setup}; \S~\ref{sec:results} presents the main results. In \S ~\ref{sec:exsistance} we present evidence for the existence of the LSC in each of the fluid layers; in \S~\ref{sec:reversal} we focus on the dynamics of LSCs, especially for cessation and reversals of LSC in each fluid layer; in \S~\ref{sec:coupling} we present flow-coupling analysis. We summarize our findings and conclude in \S~\ref{sec:conclusion}.
 
\section{Experimental setup and data analysis methods}\label{sec:setup}
    The multithermal probe technique was used to study the dynamics and coupling of the LSCs \citep{Cioni1997JFM,Eric2005PRL,hdxi2007PRE,Xie2013JFM}. The convection cell with thermistor holders is similar to the one used by \citet{Xie2013JFM}. Briefly, it is a vertical cylinder with height $H=38.4$ cm and diameter $D=19.0$ cm. The cooling and heating plates are made of $1.0$ cm thick copper with nickel coated surface. The cell was levelled to within $0.001$ rad. Two immiscible fluid layers, namely water above fluorinert FC-77 electronic liquid (3M Company, hereafter referred as FC77), with fluid height $H_{fluid}$ of each layer equals to $D$, were used as the working fluids. The aspect ratios of each of the fluid layers were thus unity. Six banks of thermistor holders located at distances $H/8$, $H/4$, $3H/8$, $5H/8$, $3H/4$, $7H/8$ from the bottom plate, with the lower three in FC77 and the upper three in water (hereafter referred to as the bottom, middle and top heights for each layers) were adhered to the sidewall. At each height these thermistor holders were distributed uniformly in eight columns around the cell periphery. A total of $48$ thermistors with accuracy of $0.01^o$C (Omega Inc., $44031$) were placed inside the holders to measure the horizontal temperature profiles with a sampling rate $0.33$ Hz. The distance between thermistor head and the fluid-sidewall contact surface was $0.7$ mm. 
  
    The measured temperature profile at a certain height was fitted to a cosine function $T_{i}=T_{0}+\delta \cos(\frac{\pi}{4} i-\theta),(i=0 \cdots 7)$ at every time step, where $T_{i}$ is the temperature reading of the $i$th thermistor, $T_{0}$ the mean temperature of the eight thermistors at one height. $\delta$ the magnitude of the cosine function, which is a measure of the LSC's flow strength, and $\theta$ is the azimuthal position where the hot plumes of the LSC go upward, which we denote as the orientation of the LSC. Hereafter, we use subscripts `W' and `77' to label quantities measured in water and FC77 respectively. The symbols $\theta_{W}^{t},~\theta_{W}^{m}$ and $\theta_{W}^{b}$ stand for LSC's orientations in water measured at top, middle and bottom heights, respectively. And $\delta_{W}^{t},~\delta_{W}^{m}$ and $\delta_{W}^{b}$ denote the respective flow strength of the LSC in water measured at these heights. The same definitions apply to symbols $\theta_{77}^{t},~\theta_{77}^{m},~\theta_{77}^{b},~\delta_{77}^{t},~\delta_{77}^{m}$ and $\delta_{77}^{b}$. The bulk temperature of water and FC77 were kept at $15^o$C and $38^o$C, respectively ($Pr_{W}=8.1$ and $Pr_{77}=19.6$). As water is a commonly used fluid, we mention only the physical properties of FC77. The density $\rho$, thermal conductivity $\kappa$ and  kinematic viscosity $\nu$ are  respectively 1739.8 kg~m$^{-3}$, 0.06 W~(mK)$^{-1}$ and $6.6\times 10^{-7}$ m$^2$~s$^{-1}$ (at temperature 38 $^o$C). The plates used are made of copper with thermal conductivity $\sim$ 400~W~(mK)$^{-1}$, which is roughly 6600 times that of FC77. By varying $\Delta T$, two sets of $Ra$, $Ra_{W}=1.23\times 10^{9}~(1.00\times 10^{9}),~Ra_{77}=1.59 \times 10^{11}~(1.33\times 10^{11})$, were achieved. As the two sets of data give statistically similar results, only those for the higher $Ra$'s will be presented below unless stated otherwise. The Reynolds number $Re=4H^2f/\nu$ based on the oscillation frequency $f$ of the LSC in water and FC77 are $1.56\times 10^3$ and $9.51\times 10^3$, respectively, which are in agreement with previous measurement in the single-layer RBC \citep{Laml2002,Xie2013JFM}. 
   
 \section{Results and discussion}\label{sec:results}         
     \subsection{Existence of the LSC}\label{sec:exsistance}

\begin{figure}
      \centerline{\includegraphics[width=\textwidth]{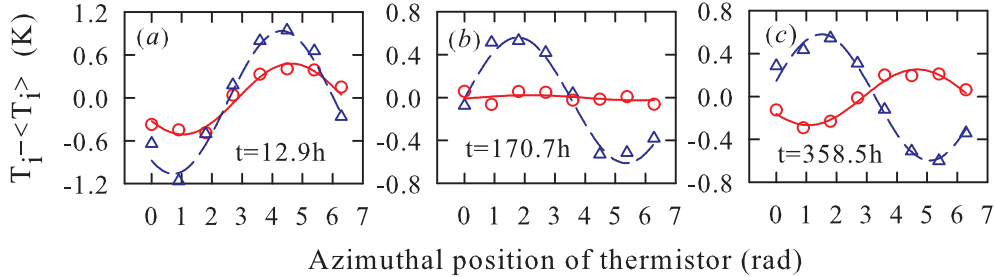}}
      \caption{Instantaneous temperature profiles measured at the mid-height in water ($\textcolor{blue}{\vartriangle}$) and FC77 ($\textcolor{red}{\circ}$) with the cosine fitting curves.}
      \label{fig:exsistance_LSC}
    \end{figure}

   The existence of the LSC is revealed by the non-zero amplitude of the first Fourier mode obtained from the horizontal temperature profiles measured along the cell periphery. Examples of instantaneous temperature profiles with cosine fitting curves measured at the mid-height of water and FC77 are shown in figure~\ref{fig:exsistance_LSC}, from which we obtain the amplitude $\delta$ and phase $\theta$ of the first Fourier mode. It is seen that in both layers, the profiles can be fitted very nicely by cosine functions. The results thus indicate that LSC exists in both fluid layers. Another evidence for the existence of the LSC is the relative weight of the first Fourier mode compared with other modes. It is found that the energy contained in the first Fourier mode is about 93.7\% of the total energy in water, and about 96.3\% of that in FC77, which also confirms the existence of the LSCs \citep{Lohse2011PoF}. In addition, the calculated cross-correlation coefficients of the orientation $\theta$ measured at different heights imply that the LSCs in both layers are of single-roll structure.
   
     A feature seen in figure~\ref{fig:exsistance_LSC} is that  $\delta_W$ is larger than $\delta_{77}$, which is due to the relatively small thermal diffusivity of FC77. Several flow states can be observed from these instantaneous profiles. One is that the azimuthal orientations of the LSCs in water and in FC77 are in phase with each other (figure~\ref{fig:exsistance_LSC}($a$)), the second is that the azimuthal orientations of the two LSCs are out-of-phase from each other (figure~\ref{fig:exsistance_LSC}($c$)) and the third one shows that the strength of LSC in FC77 almost vanishes while that in water remains above zero (figure~\ref{fig:exsistance_LSC}($b$)). 

    \subsection{Dynamics of the LSC}\label{sec:reversal}
       \subsubsection{Gross features}\label{sec:gross_feature}
       
\begin{figure}
        \centerline{\includegraphics[width=0.8\textwidth]{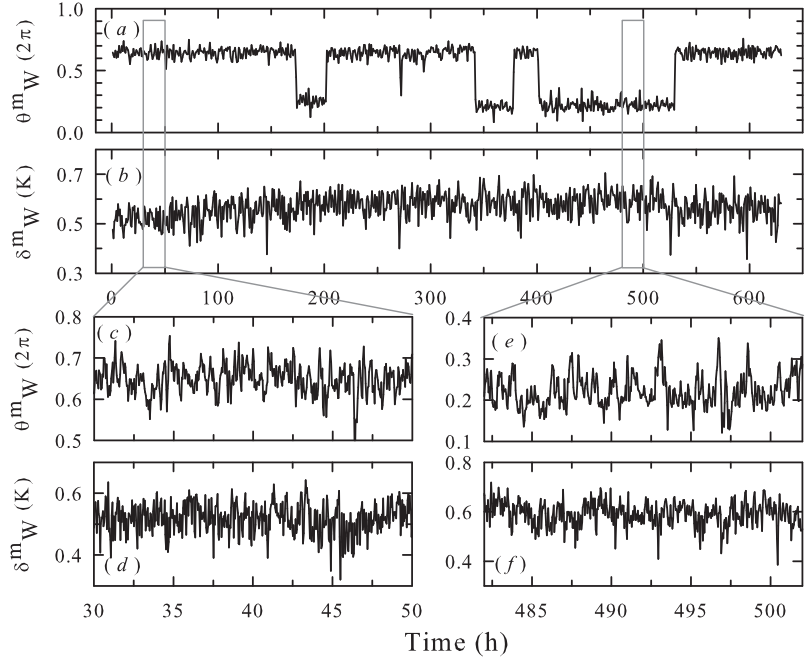}}
        \caption{The measured ($a$) flow strength and ($b$) azimuthal orientation of the LSC in water layer; ($c,d$)[($e,f$)] a segment of ~$20$ h when the orientation of the LSC is at upper [lower] plateau level. For clarity every 1 in 1000 data points are shown.}
        \label{fig:trace_water}
     \end{figure}

\begin{figure}
       \centerline{\includegraphics[width=\textwidth]{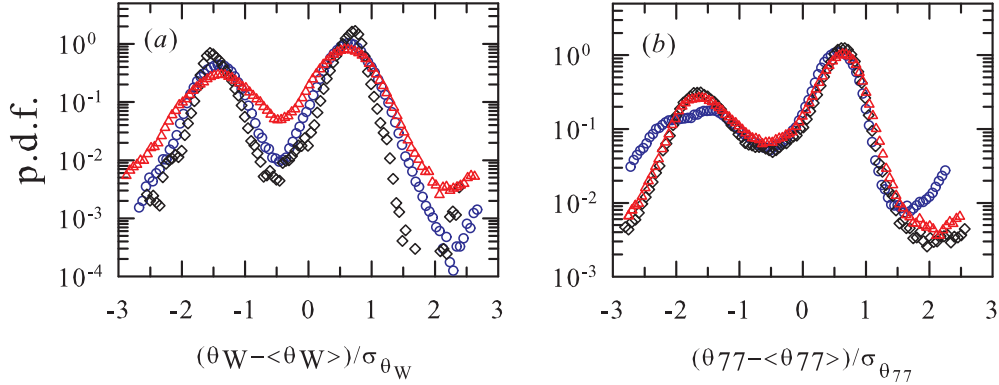}}
       \caption{Probability density function of $(a)~\theta_W$ and $(b)~\theta_{77}$ measured at the top (\textcolor{blue}{$\circ$}), middle (\textcolor{black}{$\lozenge$}) and bottom (\textcolor{red}{$\vartriangle$}) heights of the respective LSCs.}
       \label{fig:PDF_theta}
    \end{figure}

    We show in figure~\ref{fig:trace_water} the time traces of $\theta_W^m$ and $\delta_W^m$ in water. Interestingly we find that there are two
preferred orientations for the LSC, corresponding to the two plateau levels
shown in figure~\ref{fig:trace_water}($a$). The two preferred orientations are also revealed by the bimodal probability density function (p.d.f.) of normalised orientations $(\theta_W-\langle\theta_W\rangle)/\sigma_{\theta_W}$ with $\langle\theta\rangle$ as the mean value and $\sigma_{\theta_W}$ the standard deviation (figure~\ref{fig:PDF_theta}($a$)). This phenomenon contrasts sharply to that of single-layer turbulent RBC, where there is one preferred orientation for the LSC in a cylindrical cell. It was proposed that the preferred orientation in that case may be due to the Coriolis force induced by the rotation of earth \citep{Brown2006PoF}. Obviously, Coriolis force alone cannot explain the existence of two preferred orientations. We show in figure~\ref{fig:trace_water}($c,d$) and ($e,f$) a 20-hour time segment of $\theta_W^m$ and $\delta_W^m$ when the LSC in water is at two different states (corresponding to the two plateaus in figure~\ref{fig:trace_water}($a$)). From these figures, we can see that quantitatively the behaviours of $\theta_W^{m}$ and $\delta_W^m$ show no obvious difference between the two plateau levels, i.e. the orientation $\theta_W^m$ fluctuates irregularly and the flow strength $\delta_W^m$ is well above zero at all times.

       \begin{figure}
       \centerline{\includegraphics[width=0.8\textwidth]{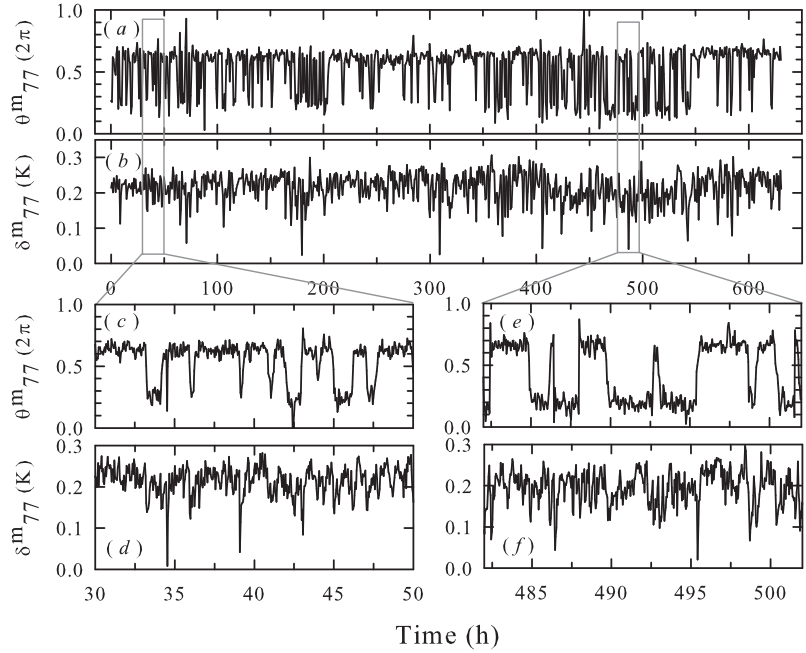}}
       \caption{The measured ($a$) flow strength and ($b$) azimuthal orientation of the LSC in FC77; ($c,d$)[($e,f$)] the same time segment of $\delta_{77}^m$ and $\theta_{77}^m$ as that in figure~\ref{fig:trace_water}($c,d$) [($e,f$)]. For clarity every 1 in 1000 data points are shown.}
       \label{fig:trace_FC}
    \end{figure}       
       
      The time traces of $\theta_{77}^m$ and $\delta_{77}^m$ in FC77 are shown in figure~\ref{fig:trace_FC}. Figure~\ref{fig:trace_FC}($a$) reveals that there are also two plateau levels in the FC77 layer, which can be seen clearly in the p.d.f.s of the normalized orientations at different height measured in that layer (figure~\ref{fig:PDF_theta}($b$)). It will be shown below that these are the same preferred orientations as those in the water layer. In figures~\ref{fig:trace_FC}($c,d$) and ($e,f$) we zoom in the time traces that have the same period as those shown in figure~\ref{fig:trace_water}($c,d$) and ($e,f$). The transitions between the two plateau levels are now obvious. This is quite different from the behaviour of the LSC in water, where there are only 6 such transitions. Another feature revealed by figure~\ref{fig:trace_FC}($d$) is that the flow strength $\delta_{77}^m$ sometimes drops to almost zero, which is not seen in the whole time trace of $\delta_W^m$. 

\begin{figure}
              \centerline{\includegraphics[width=\textwidth]{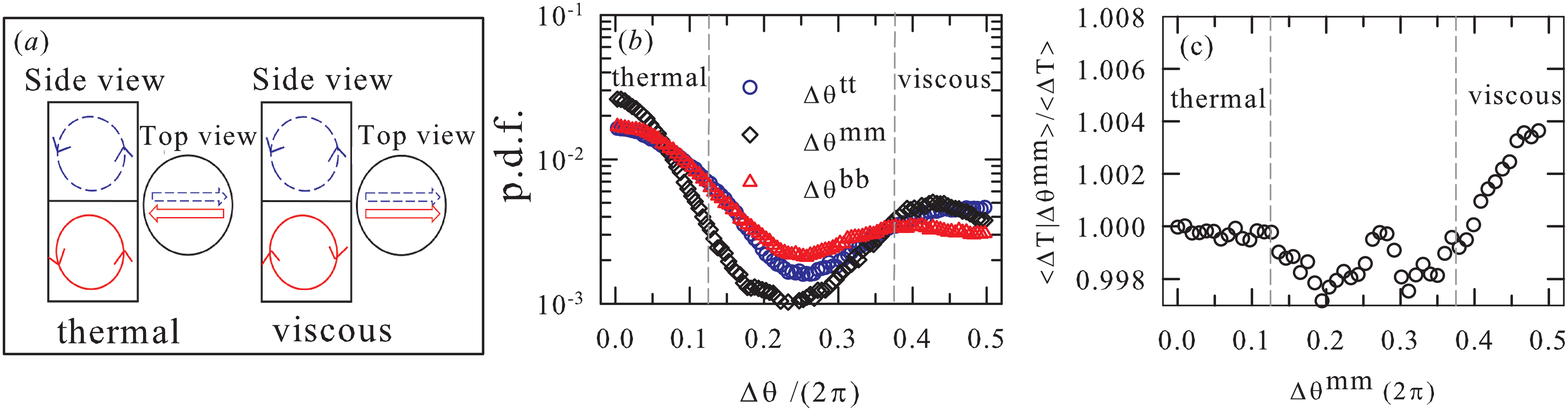}}
              \caption{($a$) Schematics of flow-coupling modes. ($b$) Probability density function of $\Delta\theta$. ($c$) Normalized temperature difference of the top and bottom plates conditioned on the flow state parameter $\Delta\theta^{mm}$. The dashed lines are the boundaries for different coupling states. Thermal and viscous stand for thermal coupling and viscous coupling, respectively.}
              \label{fig:conditional_DeltaT}
          \end{figure}

     Comparing figures~\ref{fig:trace_water}($a$) and ~\ref{fig:trace_FC}($a$), we can see that $\theta_W$ and $\theta_{77}$ will be either at the same plateau level, or one is at the upper level and the other is at the lower one. As the LSC is a coherent motion of thermal plumes erupted from the thermal boundary layer \citep{Xi2004JFM}, the interface between the two fluids may be affected by the two LSCs above and below it. This in turn may modify the behaviour of plume emissions from the interface for both the upward going warm plumes in the water and the downward going cold plumes in the FC77. This interaction leads to coupling of the two LSCs and results in their azimuthal orientations to be either parallel or anti-parallel with each other, corresponding to the `plateau levels' in the time traces of $\theta_W$ and $\theta_{77}$. These parallel or anti-parallel states of the LSC's orientation are similar to the thermal coupling and viscous coupling of the convection rolls near the onset of convection in two-layer RBC \citep{Buss2009PRE}. The same terminologies will therefore be used to define the coupling of the LSCs in the turbulent regime, which are:
       \begin{itemize}
        \item[\textit{Thermal coupling:}] the flow directions of the LSCs are the same at the      sidewall region, and are opposite to each other at the fluid-fluid interface (figure\ref{fig:conditional_DeltaT}($a$), left panel); 
   
   \item[\textit{Viscous coupling:}] the flow directions of the LSCs at the sidewall region are opposite to each other, and those at the fluid-fluid interface are the same (figure~\ref{fig:conditional_DeltaT}($a$), right panel).
      \end{itemize}
For the time periods that the LSCs are neither thermally nor viscously coupled, we call them transient states. The phase difference $\Delta\theta$ of the two LSCs are used as a flow state parameter to analyze the coupling of LSCs, which is defined as
     \begin{equation}
     \Delta\theta^{ij}=|\theta_W^i-\theta_{77}^j|\qquad (i,j=t,m,b)
     \end{equation}
   Because of the cylindrical symmetry, we reduce $\Delta\theta$ to $[0,\pi]$. For thermal coupling mode, $\Delta\theta\sim 0$, and $\Delta\theta\sim\pi$ for viscous coupling mode.

      The same preferred orientation of the two LSCs is one indication of flow-coupling of the two fluid layers. We now examine the effect of flow-coupling on the global properties of the system. The p.d.f. of the flow state parameter $\Delta\theta$ is shown in figure~\ref{fig:conditional_DeltaT}($b$). It is seen that the p.d.f.s of $\Delta\theta$ have a peak close to 0 and a small shoulder close to $\pi$, which corresponds to the thermal and viscous coupling modes respectively. The probability that the system stays in the thermally coupled mode is found to be $64\%$ and that for viscous coupling mode is only $20\%$, which suggests that the two LSCs always like to stay in parallel to each other. As the LSC in each of the fluid layer is predominantly a single roll structure, parameters $\Delta\theta^{tt},~\Delta\theta^{mm}$ and $\Delta\theta^{bb}$ essentially show similar statistics. Thus only results based on $\Delta\theta^{mm}$ will be presented in the analysis hereafter. The temperature difference $\Delta T$ across the top and bottom plates conditioned on $\Delta\theta^{mm}$ is shown in figure~\ref{fig:conditional_DeltaT}($c$). Under constant heat flux at the bottom plate, $\Delta T$ is inversely proportional to the global $Nu$ of the system. Thus, the thermal coupling mode has a higher heat transfer efficiency when compared with the viscous one. Although the heat transfer efficiency difference between the two coupling modes is small, the physics involved is clear. That is different internal flow states in turbulent thermal convection can indeed produce different global responses for the system. It was first discovered in a single-layer turbulent RBC with $\Gamma=0.5$ that the heat transport efficiency of the system depends on the internal flow modes of the LSC \citep*{CSun2005PRL,hdxi2008POF,SA2011JFM,xia2011JFM}. Finally we remark that, as a result of the interaction of the two LSCs, the interface between the two fluid layers may undergo some kind of deformation, such as surface waves. Indeed, preliminary shadowgraph observation indicates that there are wavy motions at the interface. But it is hard to ascertain whether these are surface waves or just thermal plumes travelling along the interface while being swept by the LSC. This will be a very interesting subject for future studies.   
       
 \subsubsection{Cessations and reversals}
  
    Cessation of the LSC has been well-studied in previous experiments in single-layer turbulent RBC \citep{Eric2005PRL,hdxi2006PRE,BA06JFM,hdxi2007PRE,SA2011JFM,Xie2013JFM}. In these cases  cessations are rather rare events; its frequency of occurrence is on the order of one in several days in $\Gamma=0.5$ and $1$ cells \citep{SA2011JFM,Xie2013JFM}. A strict definition of cessation requires the flow strength $\delta$ at top, middle and bottom heights of LSC to fall below a threshold value simultaneously. In order to acquire more statistics, a less strict definition has also been used, which requires only one of the above strengths to fall below the threshold. In \citet{Xie2013JFM}, this latter situation is named decoherence. In the present work, both cessations and decoherences will be studied. If after a cessation the LSC's azimuthal orientation is changed by $\pi$, it is called a cessation-led reversal. \cite{Eric2005PRL} have found that in a $\Gamma =1$ cell the LSC is equally likely to reemerge in any azimuthal orientation after a cessation, so reversals are much more rare events than cessations. In a $\Gamma=0.5$ cell, \citet{hdxi2007PRE} found that after a decoherence at the mid-height, the LSC is more likely to reverse it's direction, which is due to the complex flow mode transitions as discovered later by \citet{hdxi2008POF}. Besides the cessation-led flow reversals, the LSC sometimes undergoes a fast change of its azimuthal orientation by $\pi$ without significant reduction of it's flow strength, which is called reorientation-led flow reversal.
  
\begin{figure}
              \centerline{\includegraphics[width=\textwidth]{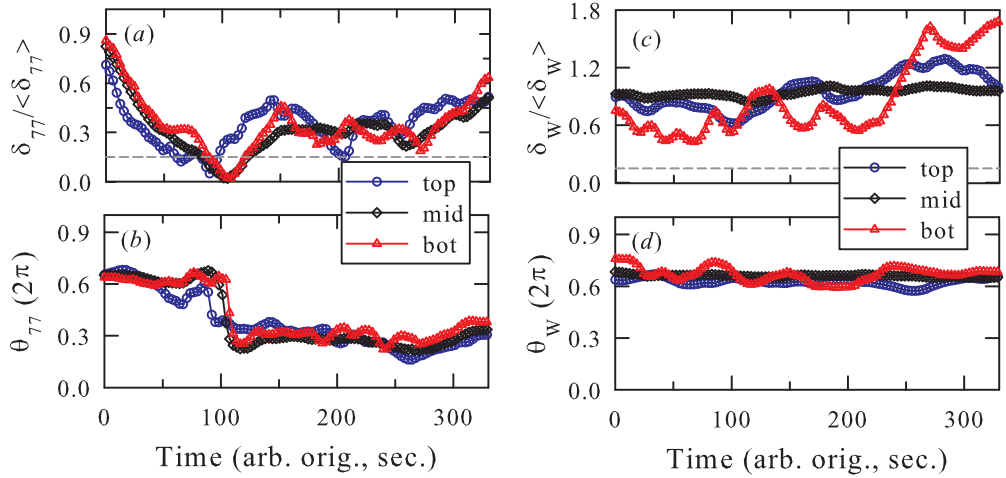}}
              \caption{An example illustrating cessation-led reversals in FC77. $\delta_{77}$ and $\theta_{77}$ are shown in $(a)$ and $(b)$, respectively. $(c)$ and $(d)$ the corresponding $\delta_W$ and $\theta_W$. The dashed lines in ($a$) and $(c)$ mark the threshold 0.15$\langle\delta\rangle$ for cessations.}
              \label{fig:cessation}
          \end{figure}

   In figure~\ref{fig:cessation} we show an example of cessations in the FC77 layer with $\delta_{77}$ measured at the three heights of the LSC in figure~\ref{fig:cessation}($a$) and $\theta_{77}$ at those heights in \ref{fig:cessation}($b$). It is seen that $\delta_{77}$ at all three heights vanish nearly simultaneously (within one LSC turnover time), meanwhile the orientation of the LSC has changed approximately by $\pi$, i.e. the LSC in FC77 has experienced a cessation and re-emerges at the opposite direction. The corresponding time trace of $\delta_W$ and $\theta_W$ are shown in figure~\ref{fig:cessation}($c$) and ($d$), where $\delta_W$ at the three heights are well above zero. It can be seen from the figures that after the cessation, the two LSCs switch from thermal coupling mode to viscous coupling mode. Operationally, a cessation is identified when $\delta$ at all three heights of the LSC are below a threshold value $A$ simultaneously (within one LSC turnover time), with $A$ ranges from 0.15$\langle\delta\rangle$  to 0.3$\langle\delta\rangle$, where $\langle\cdots\rangle$ denotes the time-averaged value. As cessations identified with different $A$ have similar statistical properties, we will use $A=0.15\langle\delta\rangle$, which is the same as in \citet{BA06JFM} and \citet{hdxi2007PRE}.
  \begin{table}
           \begin{center}
              \def~{\hphantom{0}}
                \begin{tabular}{cllllllc}
 System&$Ra$&$Pr$&Cessations&Decoherences&\tabincell{l}{Cessation-led \\reversals}&\tabincell{l}{Reorientation\\-led reversals}&Source\\[3pt]
  \vspace{2mm} 
 \tabincell{c}{77\\(two)}&$1.59\times 10^{11}$&$19.6$&1.3 (34)&\tabincell{l}{14.8~(391, t)\\ 2.6~~(61,~m)\\3.9~~(105, b)}&0.9~(24)   &16.9~(446)&$a$\\ 
\vspace{2mm}

 \tabincell{c}{W \\(two)}&$1.23\times 10^{9}$ &$8.1$&0.0 (0)&\tabincell{l}{1.4~~(38,~t)\\0.0~~(0,~m)\\22.6~(595,b)} &0.0~(0)  &0.2~(6) &$a$\\
 \vspace{2mm}
  \tabincell{c}{77 \\(single)} &$2.00\times10^{11}$&$19.4$&0.1 (3)&\tabincell{l}{1.2~(36, t)\\1.0~(30,~m)\\1.0~(28, b)}    &0.03 (1)  &0.03 (1)&$b$
        \end{tabular}
              \caption{Cessation, decoherence and reversal frequencies (day$^{-1}$) in FC77 and water, with the corresponding observed total numbers shown in brackets. `Single' and `two' refer to single-layer and two-layer RBC. The letters `t',`m' and `b' refer to top middle and bottom heights of the LSC respectively. Source: $a$ present experiment with a measurement period of 630.3 h; $b$ \citet*{Xie2013JFM} with a measurement period of 700 h.}
              \label{tab:cess}
           \end{center}
        \end{table}  
                        
\begin{figure}
         \centerline{\includegraphics[width=\textwidth]{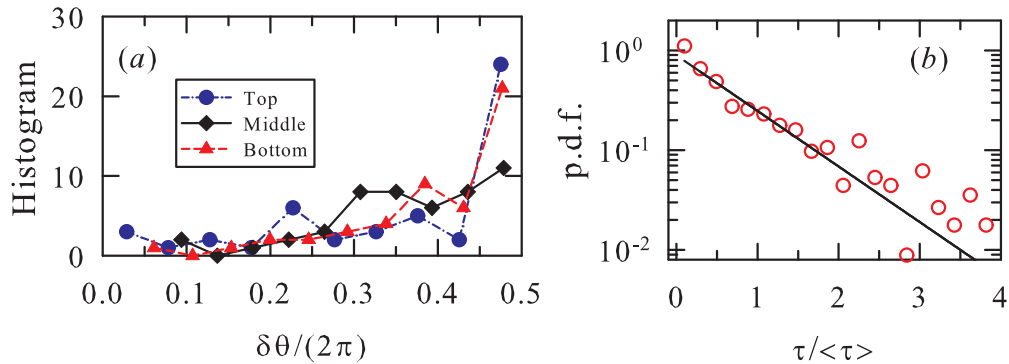}}
         \caption{($a$) Histogram of the LSC's orientation change $\delta\theta$ after a cessation in FC77. ($b$) Probability density function of time interval between reversals with solid line marking an exponential distribution.}
         \label{fig:hist_cess}
      \end{figure}

Table~\ref{tab:cess} lists the frequency (day$^{-1}$) and the total number of cessations, docoherences, cessation-led reversals and reorientation-led reversals in the present two-layer system. For reference, the same numbers for a single-layer turbulent RBC within the same parameter range are also listed. The first surprising finding is that cessation frequency ($\sim 1.3$/day) of the FC77 layer in the two-layer system is an order-of-magnitude larger than that in the single-layer case ($\sim 0.1$/day); whereas no cessation was observed in the water layer for the duration of the measurement. We note that the temperature boundary conditions for FC77 are constant heat flux at both top and bottom and those for water are constant flux at bottom and constant temperature at top. It is further found that 2/3 of the cessations will lead to flow reversals, which gives a cessation-led reversal frequency of $\sim 1$/day. This is confirmed by the histogram (figure~\ref{fig:hist_cess}($a$)) of the orientation change $\delta\theta$ before and after a cessation using data from the two $Ra$ values of the experiment. A maximum value close to $\pi$ is consistent with the fact that cessations in FC77 most likely lead to flow reversals. This contrasts sharply to the uniform distribution of $\delta\theta$ found for $\Gamma=1$ cell in single-layer turbulent RBC \citep{Eric2005PRL,hdxi2006PRE,BA06JFM}, indicating that the dynamics governing cessations are very different in the two systems. A uniform distribution implies that the system has lost its memory during the cessation and the LSC is equally likely to reemerge in any orientation.

From figure~\ref{fig:conditional_DeltaT} we see that the thermal-coupling mode has higher probability and also has slightly higher heat transfer efficiency than the viscous-coupling mode. This suggests that the two LSCs will be more likely to couple thermally after a cessation. Indeed, the system is found to be thermally coupled $3/5$ of the time after a cessation. In addition, it is found that if the system is thermally-coupled before a cessation, it can be either thermally or viscously coupled after the cessation. But if it is viscously-coupled before a cessation, it will always end up thermally coupled.  
      
We note that in a single-layer RBC experiment, \cite{BA06JFM} have reported a cessation frequency of $\sim1.5/$day. As some of the cells used in their measurements had thermistors only at the mid-height, it is clear that, with the present classification, many of the cessations found in their experiment should be counted as decoherences here. The decoherence rate at mid-height  is $\sim2.6/$day for the FC77 layer in the present case and $\sim1.0/$day for the single-layer FC77 \citep{Xie2013JFM}. It is seen that these numbers are in general consistent with decoherence frequency found by \cite{BA06JFM}. It is also consistent with the decoherence frequency of $\sim1.0/$day found by \cite{hdxi2006PRE}. Table~\ref{tab:cess} also shows that a large number of decoherences are observed at the bottom of the water and top of the FC77, suggesting that the interaction of the two convective flows through the liquid interface decreases the stability of the LSCs. The fact that the bottom of water layer and top of FC77 are less stable can also be seen from the p.d.f.s of $\theta^{b}_{W}$ and $\theta^{t}_{77}$ shown in figure~\ref{fig:PDF_theta}. 

The reorientation-led reversals is seen to have increased well over two orders of magnitude compared to the single-layer case with similar $Ra$ and $Pr$. The corresponding reversal frequency is $\sim 17$/day. Figure~\ref{fig:hist_cess}($b$) plots the p.d.f. of the time interval between adjacent reversals including both cessation-led and reorientation led one. The solid line is an an exponential fit to the data, which indicates that the reversals in FC77 are a Possion process, i.e. the occurrence of successive reversals are independent of each other. This suggests that cessations/reversals and reorientations are stochastic processes. 

 \begin{figure}
      \centerline{\includegraphics[width=0.9\textwidth]{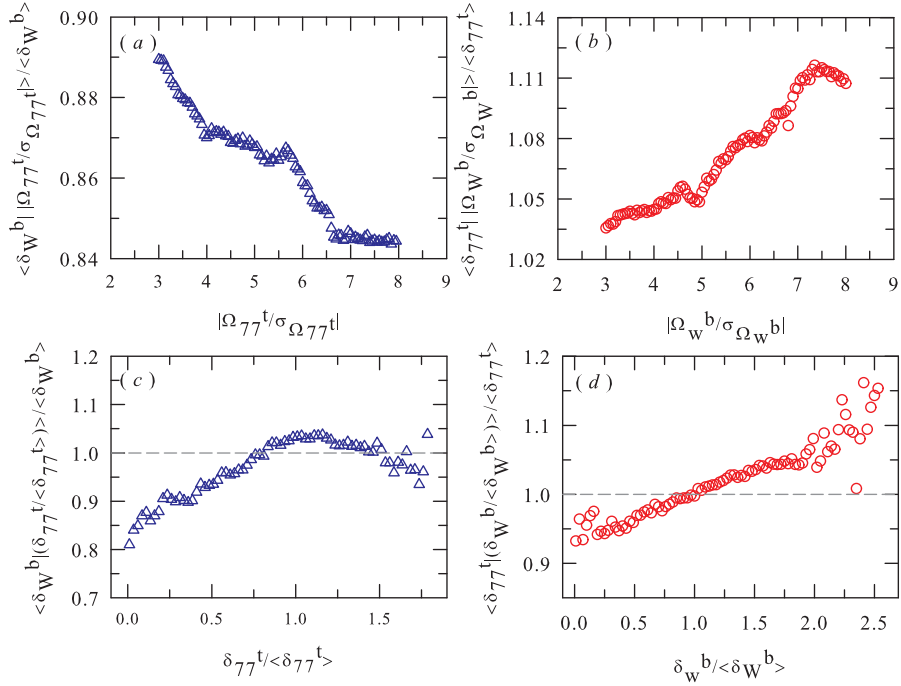}}
      \caption{($a$) Normalized flow strength in water conditioned on angular speed of LSC in FC77; ($b$) Normalized flow strength in FC77 conditioned on angular speed of LSC in water; ($c$) Normalized flow strength in water conditioned on flow strength of LSC in FC77; ($d$) Normalized flow strength in FC77 conditioned on flow strength of LSC in water.}
      \label{fig:A_condition_A}
   \end{figure}

We now discuss these phenomena in the two-layer system. The coupling of the two LSCs may be understood as a system with two potential wells. As the potential energy of the thermal coupling is lower than that of viscous one (the p.d.f. shown in figure~\ref{fig:conditional_DeltaT}($b$) may be viewed as an inverted potential energy), the LSCs are more likely to couple thermally. However, turbulent fluctuations will kick the system out of this state from time to time  either via cessations/decoherences or (large) reorientations. When it became viscously coupled as a result of such an event, the system is less stable. It will then try to go back to the thermally-coupled mode as long as turbulent fluctuations are large enough to carry it across the barrier.  As in most of these instances the LSC in the water layer remains unchanged, when the LSC in the FC77 layer reestablishes itself (in the case of cessations) it will prefer to align either parallel or antiparallel with the LSC in the water layer, i.e. to be either thermally- or viscously-coupled, but with a higher probability for the former. This is what led to the non-uniform distribution of the orientation change ($\delta\theta$) after a cessation. Comparing to the uniform distribution of $\delta\theta$ in the single-layer case, we see that this symmetry-breaking is imposed by the orientation of the LSC in the water layer. Therefore, the dynamics of cessations and reversals of the LSC in the two-layer system is a result of the coupling of the two layers, which is very different in comparison with the single-layer system. Note that the above physical picture indicates that the average lifetime of the thermal coupling mode should be longer than that of the viscous one. Indeed, we find that the average lifetime of the thermal coupling sate is $59.2 \tau_{77}$ and that of the viscous one is $17.0 \tau_{77}$, where $\tau_{77}=24.10$ s is the LSC turnover time.

   \subsection{Flow coupling}\label{sec:coupling}
    
    In this section, we investigate the influence of the LSC in water (FC77) on that of FC77 (water). The average flow strength $\delta_{W}^{b}$ in water conditioned on the angular speed $\Omega_{77}^t$ of the LSC in FC77 is shown in figure~\ref{fig:A_condition_A}($a$), where $\delta_{W}^{b}$ is normalized by its time averaged value $\langle\delta_W^b\rangle$, and $\Omega_{77}^t$ is normalized by its standard deviation $\sigma_{\Omega_{77}^t}$. It is clear that the flow strength in water decreases as the angular speed of the LSC in FC77 increases. In other words, the flow strength in water \textit{weakens}, when the LSC in FC77 rotates faster azimuthally. Similarly, we plot in figure~\ref{fig:A_condition_A}($b$) the averaged flow strength $\delta_{77}^{t}$ conditioned on the angular speed $\Omega_W^b$. We find that when the LSC in water rotates faster azimuthally, the flow strength in FC77 becomes \textit{stronger}, which is opposite to the situation shown in figure~\ref{fig:A_condition_A}($a$). The different behaviour suggests that the effects of the LSCs in the two layers on each other are not symmetric. The normalized flow strength of water conditioned on the flow strength of FC77 is shown in figures~\ref{fig:A_condition_A}($c$) and that of FC77 conditioned on the flow strength of water is shown in figure~\ref{fig:A_condition_A}($d$). It is seen from the figures that when the strength of one LSC increases, that of the other also increases. 
            
\section{Conclusion}\label{sec:conclusion}
In this study we have made an experimental investigation of the dynamics and flow-coupling in turbulent Rayleigh-B\'enard convection with two immiscible liquids. The two layers, water above FC77, each has an aspect ratio of unity. It is found that there is one large-scale circulation (LSC) in each of the layers and they span the height of that layer. The dynamics of LSCs in the two layers are quite different from that in single-layer turbulent thermal convection in the same parameter range. The major findings are ($1$) there are two preferred orientations for the LSCs in both water and FC77, rather than one preferred orientation in the single-layer case. It is shown that these preferred orientations correspond to two coupling modes of the LSCs, the thermal coupling mode where flows are opposite to each other above and below the interface of the two fluids and the viscous coupling mode where flows above and below the interface have the same direction. Furthermore, the thermal coupling mode is found to be the predominant one and it also has slightly higher heat transfer efficiency for the whole system. ($2$) Cessations in FC77 of the two-layer system occur an order-of-magnitude more frequently than they do in the single-layer case. ($3$) Cessations most likely lead to flow reversals in FC77. This property contrasts sharply with the uniform distribution of the azimuthal angular change of the LSC found in single-layer turbulent RBC. ($4$) Flow reversals resulting from cessations and reorientations in the FC77 layer occur well-over two orders of magnitude more frequently than they do in the single-layer case. The greatly enhanced reversals may be understood as transitions of the system between its two metastable states, the thermal and viscous coupling modes. The fact that most cessations (in the FC77 layer) end up as reversals can also be understood as a symmetry breaking imposed by the orientation of the LSC in the water layer, which remained unchanged most of the time. It is further found that the influences of LSCs to each other are not symmetric: the flow strength in water becomes weaker when the LSC in FC77 rotates faster azimuthally, whereas the flow strength in FC77 becomes stronger when the LSC in water rotates faster azimuthally. 

Results presented in this paper may be of some relevance to flow reversals in coupled flow systems, such as the polarity reversals in the convective outer core of the earth which are coupled to the lower mantle, or wind reversals in the atmosphere which are also coupled with the ocean current. Additionally, it is known that reversals in 3D systems occur much less frequently than they do in 2D convection systems and the properties and mechanisms of flow reversals in 3D and 2D systems are different
\citep{RNiReversal2D}. Thus, with its greatly enhanced reversal/cessation frequency, the FC77 layer  in the two-layer system may be used as a platform for the study of flow reversals/cessations in 3D turbulent convection. While the exact reason(s) for the much-different behaviors of the FC77 and the water layers remain unknown, it is noted that the former has temperature boundary condition of constant heat flux at both its top and bottom, whereas the latter has constant flux at its bottom and constant temperature at its top boundary.

\section*{Acknowledgements}
    We thank P. Wei, R. Ni and S.-D. Huang for discussions. This work was supported by the Hong Kong Research Grants Council (RGC) under grant No.CUHK 403811.


\bibliographystyle{jfm}

\end{document}